\def\erot {\dot E}

\def\lum{erg~s$^{-1}$}

\def\norm{ph~cm$^{-2}$~s$^{-1}$~TeV$^{-1}$}
\def\diff{cm$^{2}$~s$^{-1}$}

\newcommand{\rev}[1]{\textcolor{black}{#1}}

\documentclass[a4paper,11pt]{article}
\usepackage{pos}
\usepackage{journalnames}
\usepackage{color}

\title{Unveiling TeV halos among unidentified extended TeV sources}

\author[a,b]{Michela Rigoselli}
\author[c,a,d,e]{Sarah Recchia}
\author[f,g,a]{Alberto Bonollo}
\author[b]{Silvia Crestan}
\author[c]{Giada Peron}
\author[b]{Andrea Giuliani}
\author[b]{Sandro Mereghetti}

\affiliation[a]{INAF -- OA Brera,  Via Brera 28, I-20121 Milano, Italy}
\affiliation[b]{INAF -- IASF Milano,  Via Alfonso Corti 12, I-20133 Milano, Italy}
\affiliation[c]{INAF -- OA Arcetri, Largo Enrico Fermi
5, I-50125, Firenze, Italy}
\affiliation[d]{Dipartimento di Fisica, Università di Torino Via P. Giuria 1, I-10125 Torino, Italy}
\affiliation[e]{INFN Torino Via P. Giuria 1, I-10125 Torino, Italy}
\affiliation[f]{IUSS Pavia, Palazzo del Broletto, piazza della Vittoria 15, I-27100 Pavia, Italy}
\affiliation[g]{Dipartimento di Fisica, Università di Trento, Via Sommarive 14, I-38123 Trento, Italy}

\emailAdd{michela.rigoselli@inaf.it, sarah.recchia@inaf.it}

\abstract{In recent years, the number of known sources emitting very- and ultra-high-energy gamma-rays has increased significantly thanks to facilities such as LHAASO and HAWC. Many of the observed sources are still unidentified or poorly constrained due to the limited angular resolution of these instruments; however, it is now ascertained that approximately half of them have a pulsar in coincidence.

Some of these unidentified extended sources may be the result of the diffusion of leptons accelerated by the pulsar itself or in its nebula to energies exceeding 50 TeV. This new class of sources, called TeV halos, is characterized by a peculiar radial profile that, if properly resolved, is key to distinguishing them from other TeV sources that are associated with a pulsar, such as supernova remnants and pulsar wind nebulae.

In this contribution, we consider all the pulsars which are spatially coincident with an unidentified extended TeV source, in order to quantify whether its spin-down power, age and distance allow the pulsar to produce a TeV halo with the observed flux and extension.

We also investigate how the next generation of Imaging Atmospheric Cherenkov Telescopes (IACTs), namely the Cherenkov Telescope Array Observatory (CTAO) and the ASTRI Mini-Array, will observe and characterize these TeV halos. We present a set of simulated sources with the expected morphology and spectrum, and we show for which of them we can distinguish between TeV halos and other classes of extended sources.

This research has made use of the CTA Simulation Telescope Models (CTA South with 14 Medium-Sized Telescopes and 37 Small-Sized Telescopes, CTA North with 4 Large-Sized Telescopes and 9 Medium-Sized Telescopes; version prod5 v1.0) and the CTA instrument response functions (version prod5 v1.0) provided by the CTA Observatory and Consortium, as well as the ASTRI Mini-Array (9 Telescopes) instrument response functions provided by the ASTRI Project.}

\ConferenceLogo{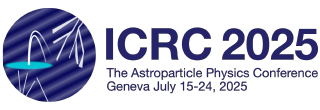}

\FullConference{39th International Cosmic Ray Conference (ICRC2025)\\
 15–24 July 2025\\
Geneva, Switzerland\\}

\begin{document}
\maketitle
\section{Introduction}
TeV halos (also referred to as pulsar halos) are a new category of Galactic Very-High-Energy (VHE; $\geq 100$ GeV) $\gamma$-ray sources. They are sources extended for tens of parsec around middle-aged, isolated pulsars.
The first TeV halos to be discovered were the Geminga and the Monogem halos, that soon \rev{became} the prototypes for this class \citep{2017Sci...358..911A}. They are powered by the two middle-aged ($\tau\sim 10^5$ yr), relatively powerful ($\erot \sim 10^{34}$  \lum), and nearby ($d\sim250$ \rev{pc}) pulsars PSR J0633$+$1746 and PSR B0656$+$14.

The extension may be explained by relativistic electrons up-scattering the Cosmic Microwave Background (CMB) in a region with a diffusion coefficient 2--3 orders of magnitude lower than the average one inferred for the interstellar medium (ISM) $D({100\, \rm TeV}) \gtrsim 10^{29}$ \diff\ 
\citep{2017PhRvD..96j3016L,2020PhRvD.101j3035D}. 
This process gives rise to halos with a peculiar, fastly decreasing radial profile. However, the formation mechanism of TeV halos remains largely unknown \citep{2022NatAs...6..199L}.

Numerous efforts have been made to detect counterparts to TeV halos, especially in the X-ray \rev{and $\gamma$-ray} bands \citep[e.g.][]{2024A&A...689A.326M,2024A&A...683A.180K,2025arXiv250402185A,2025arXiv250408689G} and in the radio \citep[e.g.][]{2024arXiv240506739H}, but no significant detections were found. \citep{2025arXiv250117046N} detected excess X-ray emission in the Monogem Ring with a very soft spectrum ($E^{-3.7}$) and a morphology that is much more compact than the TeV emitting region, suggesting a spatially varying magnetic field.

In recent years, some other halo candidates have been identified thanks to the larger sky coverage and increased sensitivity at TeV energies of Extensive Air Shower (EAS) arrays, such as HAWC \citep{2020ApJ...905...76A} and LHAASO \citep{2024ApJS..271...25C}. The most promising ones are those powered by PSR J0622$+$3749 \citep{2021PhRvL.126x1103A}, PSR J0359$+$5414 \citep{2023ApJ...944L..29A}, and PSR J0248$+$6021 \citep{2025SCPMA..6879504C}, but many others are potentially hidden in the long list of unidentified sources, most of which are in spatial coincidence with a middle-aged pulsar \citep[e.g.][]{2017ATel10941....1R,2018ATel12013....1B, 2024ApJ...968..117Z}. Those sources are poorly constrained due to the limited angular resolution of EAS arrays. The next generation of Imaging Atmospheric Cherenkov Telescopes (IACTs), namely the Cherenkov Telescope Array Observatory (CTAO) \citep{ctao} and the ASTRI Mini-Array \citep{astri}, will provide the necessary spatial resolution (few arcmin) to properly resolve these sources and distinguish between TeV halos and other classes of extended sources.

In this contribution, we consider all the pulsars which are spatially coincident with unidentified extended TeV sources (Section~\ref{sec:pulsar}), and we design a set of proper spatial and spectral models for possible configurations of a TeV halo around these pulsars (Section~\ref{sec:model}). We investigate for the specific case of PSR J2238+5903, through simulations based on these models, the prospects for observations of such spatial structures with the ASTRI Mini-Array and CTAO (Section~\ref{sec:sim}).

\section{Pulsars selection}
\label{sec:pulsar}

\begin{figure*}
  \centering \includegraphics[width=0.667\columnwidth]{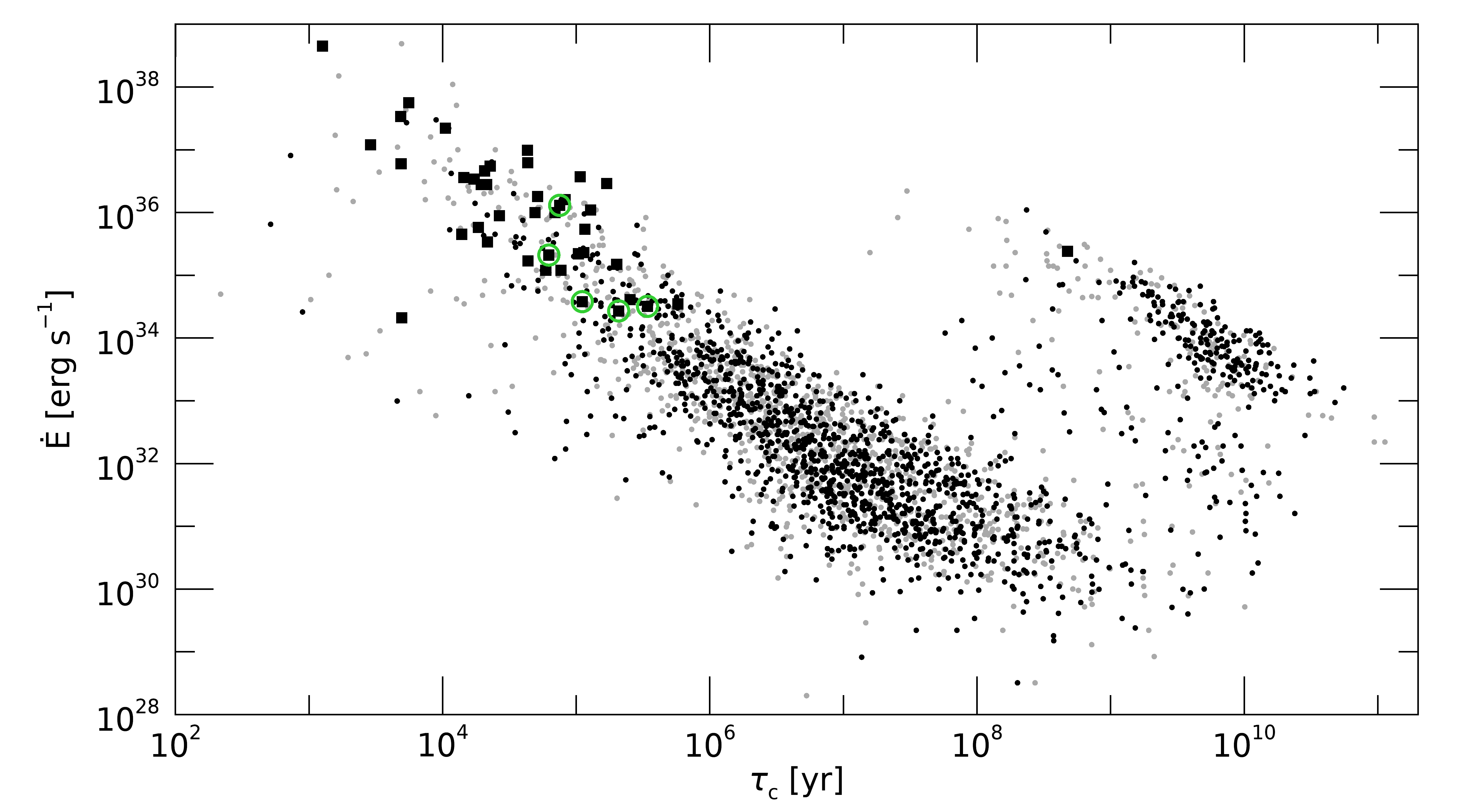}
  \caption{\footnotesize Rate of rotational energy loss $\erot$ VS characteristic age $\tau_c$ of all the pulsars in the ATNF catalog (grey dots), all the pulsars visible by LHAASO (black dots), all the pulsars in spatial coincidence with a LHAASO source (black squares), and the confirmed/candidate TeV halos (green open circles).}
 \label{fig:pulsars}
\end{figure*}

Fig.~\ref{fig:pulsars} shows the spin-down luminosity $\erot$ as a function of the characteristic age $\tau_c$ of all the pulsars of the ATNF catalog \citep{2005AJ....129.1993M}. The black squares show the 41 pulsars which are in spatial coincidence with LHAASO and HAWC sources. \rev{The Geminga and Monogem halos, plus the aforementioned most promising candidates,} are highlighted with green open circles. We note that all the TeV sources correspond to pulsars having $\erot>10^{34}$ \lum\ and $\tau<10^6$ yr, with the exception of the millisecond pulsar (MSP) PSR J0218$+$4232. This MSP is in a close binary system with a white dwarf \citep{2021ApJ...922..251A}, but its association with a TeV source is yet to be confirmed. In fact, the closest LHAASO sources, 1LHAASO J0206+4302u, J0212+4254u, and J0216+4237u
concur to create the enigmatic ``peanut'' source, whose origin is still unknown \citep{2024arXiv240702478B,2024arXiv240802070X}.
Moreover, the survey conducted by \citep{2025PhRvD.111d3014A} on HAWC data found no halo candidates among MSPs,  despite such old pulsars can have $\erot$ up to $10^{36}$ \lum\ and models have suggested that electrons may be accelerated in their magnetospheres \citep{2021ApJ...923..194H} or intrabinary shocks \citep{2020ApJ...904...91V} up to to tens of TeV energies or even more..

\section{TeV halo model}
\label{sec:model}
Given a certain particle spectrum at the pulsar, the spatial distribution of escaped particles that form the TeV halo around it, will depend only on the age or energy losses, on the spin-down power of the pulsar, and on the diffusion coefficient around it. 
We assume that cosmic ray (CR) leptons, released by a pulsar of age $\tau_{\rm age}$ and current spin-down luminosity $\dot{E}$, with power-law spectrum $\propto E_e^{-\alpha_e}$, diffuse isotropically with coefficient $D(E_{e}) = D_{100} E_{e 100}^{\delta}$ and lose energy via inverse Compton scattering (ICS) on the CMB and synchrotron  over a timescale $\tau_L(E_{e 100})$, with energy normalized to 100 TeV.
The loss time $\tau_L$, also taking into account the Klein-Nishina regime on the CMB, can be approximated as Eq.\ S7 of \citep{2017Sci...358..911A},
\begin{equation}
    \tau_L(E_{e 100}, B_{\mu G}) \approx \frac{3.3\times 10^3}{E_{e 100}} \frac{1}{0.024 B_{\rm \mu G}^2 + \frac{0.26}{(1+0.97\, E_{e 100})^{3/2}}} \; \rm yr,
\end{equation}
which corresponds to $\sim 10^4\, \rm yr$ for 100 TeV electrons when $B= 3\, \rm \mu G $.

\rev{In the assumption of constant spin-down luminosity,} the lepton distribution at distance $r$ can be approximated as \cite{1995PhRvD..52.3265A}
\begin{equation}\label{eq:fe}
    f_e(E_e, r) \approx \frac{\xi_{\rm CR}\,q_{e}^{*} (E_e/E_e^*)^{-\alpha_e}}{4\pi D(E_e) r} {\rm erfc} \left(\frac{r}{r_d(E_e)}\right),
\end{equation}
where $\tau_{\rm eff} = \text{min}(\tau_L, \tau_{\rm age})$ and
\begin{equation}\label{eq:rd} r_d (E_e) \equiv 2\sqrt{D(E_e) \tau_{\rm eff}(E_e)} \approx 26~ 
\left( \frac{D_{100}}{5\times10^{27} ~\rm cm^{2}~s^{-1}} \right)^{1/2} 
\left( \frac{\tau_{\rm eff}}{10^4~\rm yr} \right)^{1/2}
~\rm pc
\end{equation}
is the diffusion radius. 
The latter sets how far particles travel during a timescale $\tau_{\rm eff}$ and, consequently, the spatial extension of the emission. Notice that this expression is valid if the spin-down luminosity remains roughly constant during $\tau_{\rm eff}$. Since in this work we consider photons with $E_{\gamma}\geq 1~\rm TeV $, produced by leptons of $E_e \gtrsim 10~\rm TeV$, and sources with characteristic age $\tau_{\rm age} \approx \text{a few}\;  10^4~\rm yr$, the approximation is expected to be reasonably accurate.

The normalization $q_e^{*}$ is computed for total conversion of $\dot{E}$ into CRs, while $\xi_{\rm CR}$ is the acceleration efficiency. 
Given the relation between the mean electron energy $\tilde{E}_e(E_{\gamma})$ and $\gamma$-ray energy $E_\gamma$ in the ICS process (see Eq.~S12 of \citep{2017Sci...358..911A}),  photons of energy $E_{\gamma}^*$ are mainly produced by electrons of energy $E_e^{*} = \tilde{E}_e(E_{\gamma}^*)$. The $\gamma$-ray surface brightness (SB) corresponding to the CR distribution in Eq.~\ref{eq:fe}, at photon energy $E_{\gamma}$  and projected distance $r$, can be approximated as follows:
\begin{equation}\label{eq:SB}
    {\rm SB}(E_{\gamma}, r) = G(E_{\gamma})\times F(E_{\gamma}, r),
\end{equation}
where
\begin{equation}
    G(E_{\gamma}) = \xi_{\rm CR}\;   \frac{Q_{\gamma}^*\, (E_{\gamma}/E_{\gamma}^*)^{-\alpha_{\gamma}}}{(2\pi)^2\, r_d^2(\tilde{E}_e)}.
\end{equation}
The expression can be interpreted as follows: given the source injection $q_e (E_e) = q_e^* (E_e/E_e^*)^{-\alpha_e}$,  the total (integrated in space) $\gamma$-ray emissivity is 
$Q_{\gamma}(E_{\gamma}) E_{\gamma}\, d E_{\gamma} \approx q_e(\tilde{E}_e)\, \tilde{E}_e\, d \tilde{E}_e  \, \frac{\tau_{\rm eff}(\tilde{E}_e)}{\tau_{\rm ICS}(\tilde{E}_e)}$, where $\tau_{\rm ICS} = \tau_L(B_{\rm \mu G} = 0)$ is the loss time associated to ICS only.
In the calorimetric regime suitable for the present case, only a fraction $\tau_{\rm eff}/\tau_{\rm ICS}$ of the lepton energy is converted into ICS photons and:
\begin{equation}
Q_{\gamma}^* \approx q_e^* \left(\frac{E_e^*}{E_{\gamma}^*} \right) \left| \frac{d \tilde{E}_{e}}{d E_{\gamma}} \right|_{E_{\gamma}^*}\frac{\tau_{\rm eff}(E_e^*)}{\tau_{\rm ICS}(E_e^*)}.
\end{equation}
Using the same relation between $Q_{\gamma}$ and $q_e$, the slope of the electron injection spectrum, $\alpha_{e}$, can be estimated from the  
slope of the $\gamma$-ray emission, $\alpha_{\gamma}$, taking into account the  dependence $\tilde{E}_e \propto E_{\gamma}^{a(E_{\gamma})} $ appearing in Eq.~S12 of \citep{2017Sci...358..911A} and approximating all the terms as power laws around a pivot $\gamma$-ray energy $E_{\gamma}^*$. 

For $F(E_{\gamma}, r)$, it describes the spatial morphology of the emission derived from integration of the spatial part of Eq.~\ref{eq:fe} along the line of sight (los):
\begin{equation}
    F(E_{\gamma}, r) \equiv \int_{\rm los} \frac{r_d(\tilde{E}_e)}{r} {\rm erfc} \left(\frac{r}{r_d} \right) \frac{ds}{r_d} \approx \frac{0.74\,r_d}{r+0.08\, r_d}\,e^{-\left(\frac{r}{r_d}\right)^2},
\end{equation}
where we remark that $r_d$ is a function of $\tilde{E}_e(E_{\gamma})$, and that the last approximate equality results from fitting numerically integrated spatial profiles.

\begin{figure*}
  \centering
  \includegraphics[width=0.49\columnwidth]{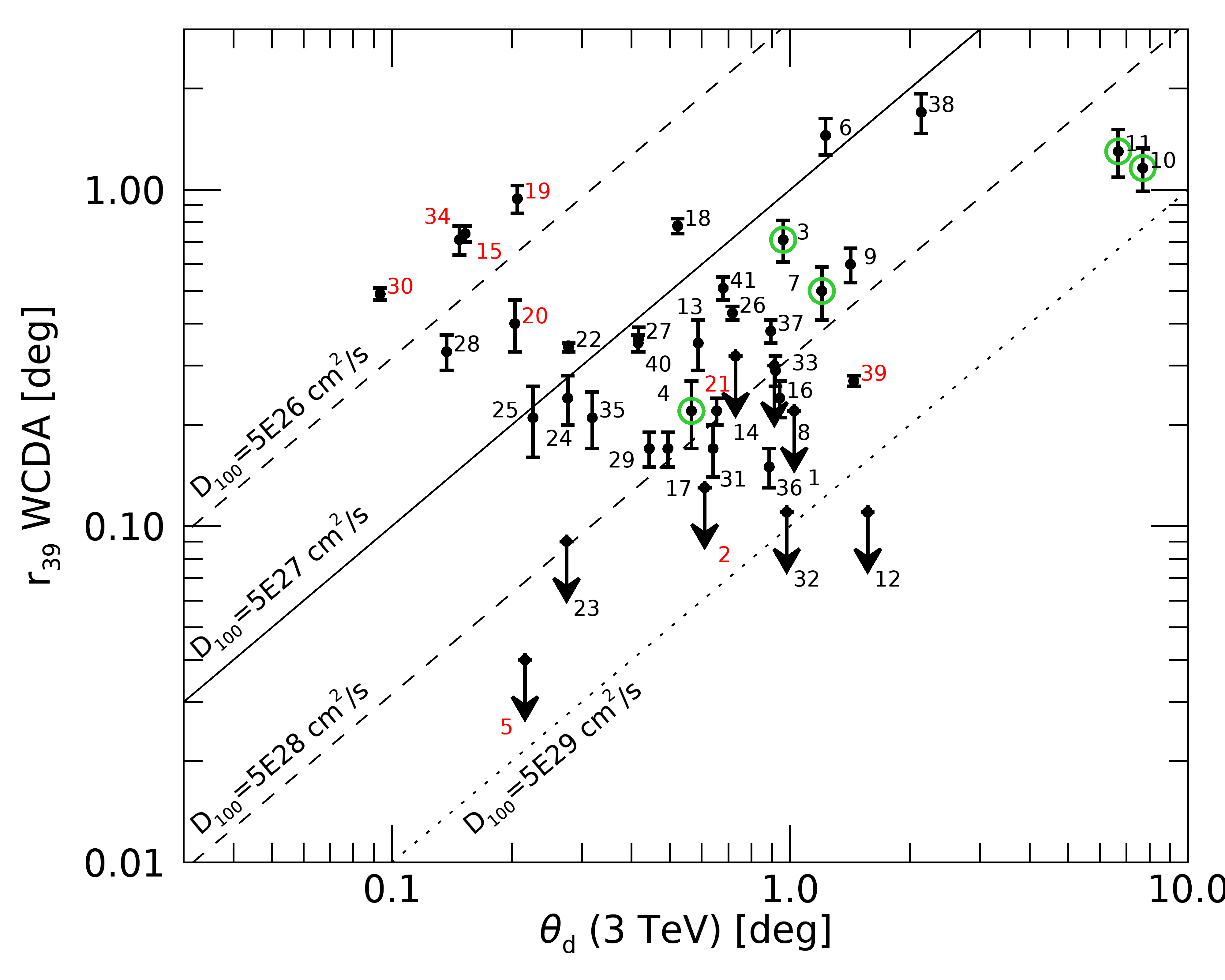}
  \includegraphics[width=0.49\columnwidth]{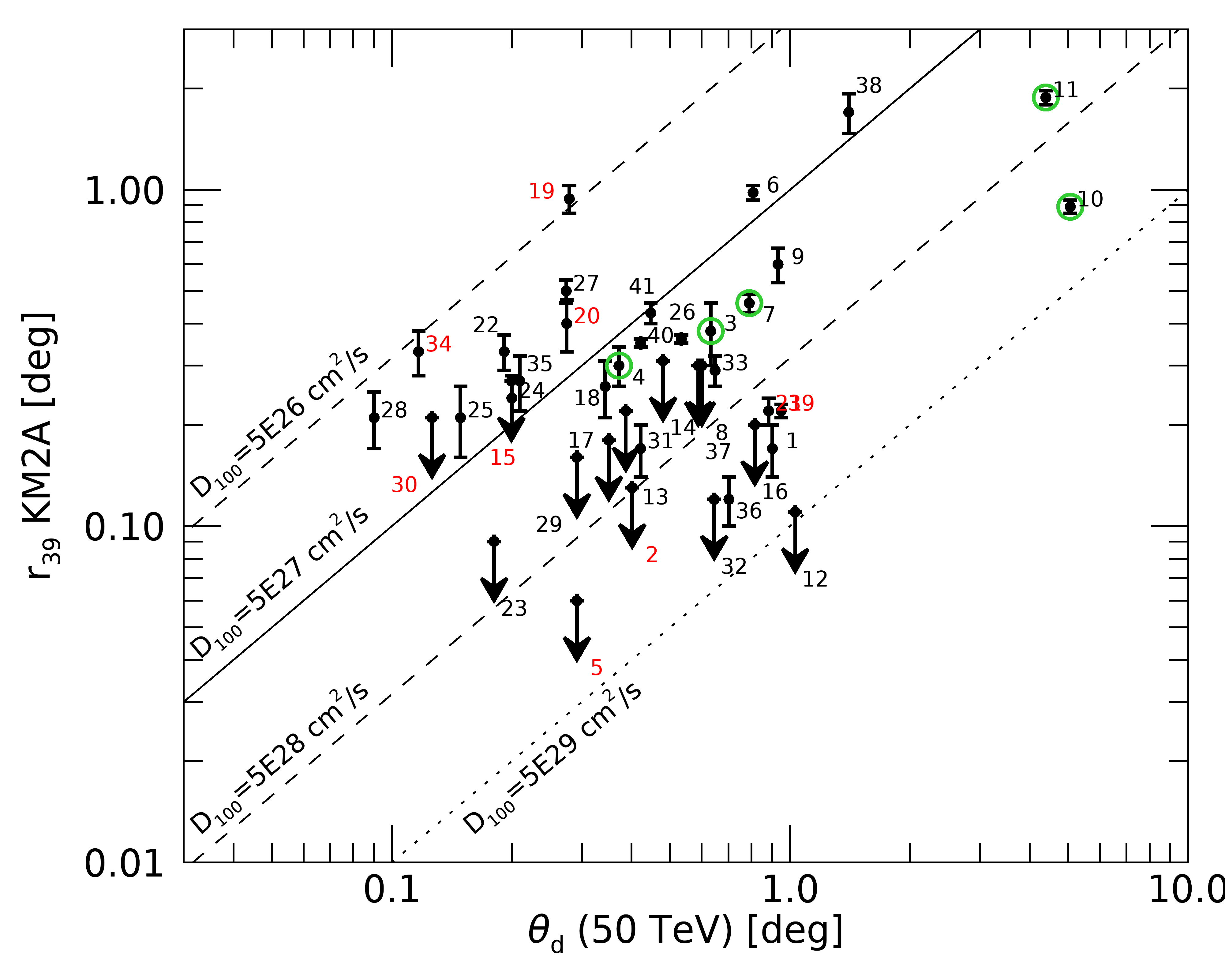}
  \caption{\footnotesize
 $\theta_d$ computed at $E_\gamma=3$ TeV (left panel) and $E_\gamma=50$ TeV (right panel) versus the extension of the corresponding LHAASO-WCDA and LHAASO-KM2A sources.
  The solid line represents $r_{39}=\theta_d$, while the two dashed lines represent the fiducial range of $\theta_d$, computed for $D_{100} = 5\times 10^{26}$ and $5\times 10^{28}$ \diff. The dotted line represents $\theta_d$ computed for a typical Galactic diffusion coefficient of $5\times 10^{29}$ \diff.
  In both panels the excluded sources (see text) are numbered in red, the confirmed/candidate TeV halos are highlighted by green open circles.
  }
 \label{fig:radii}
\end{figure*}

For each of the 41 pulsars matching LHAASO sources, we computed the expected $\theta_d = {\rm atan} (r_d/d)$, being $d$ the nominal pulsar distance. 
We assumed $D_{100}=5\times10^{27}$ \diff, $B=3$ $\mu G$,  $\delta=1/3$, in line with the HAWC analysis of Geminga \citep{2017Sci...358..911A}. 
Fig.\ \ref{fig:radii} shows $\theta_d$ computed at $E_\gamma=3$ TeV (left panel) and $E_\gamma=50$ TeV (right panel) versus the extension of the corresponding LHAASO-WCDA and LHAASO-KM2A sources in terms of $r_{39}$\footnote{\rev{$r_{39}$ is the 39\% containment radius of the best-fitting 2D-Gaussian model, corresponding to 1$\sigma$.}} \citep{2024ApJS..271...25C}.
Due to the dependence of $\theta_d$ on the pulsar and on the medium parameters, the diffusion coefficient is the primary source of uncertainty. Assuming that it can vary from 0.1 to 10 times the adopted value, we estimated $\theta_d$ to vary within a factor $0.1^{1/2}-10^{1/2}\sim 0.316-3.16$ (dashed lines). In the Figure we also show $\theta_d$ computed for a typical Galactic diffusion coefficient of $5\times 10^{29}$ \diff\ (dotted line).

We restricted our sample to 32 candidate halos, considering both \rev{the overall properties of the sources and the spatial extents of the putative associated pulsar halos}, as illustrated in Fig.\ \ref{fig:radii} (the pulsars have been numbered in alphabetical order). These 32 sources exclude young pulsars (i.e.\ number 5, 15, 19, 20, 21 and 30, all of which have $\tau_c < 10^4$~yr), since the TeV halo emission model adopted in this work can be applied to middle-aged pulsars only, and the aforementioned MSP J0218+4232 (number 2). 
Finally, we excluded pulsars (i.e.\ number 34 and 39) in spatial coincidence with doubtful LHAASO sources, which probably \rev{result from} the contribution of multiple, unresolved sources.

The chosen sample comprises all known (i.e.\ number 10, 11) and possible (number 3, 4, 7 are the most promising candidates) TeV halos, plus sources that could be supernova remnant (SNR) or pulsar wind nebula (PWN) powered.
The comparison between expected and observed properties of these sources shown in Fig.\ \ref{fig:radii} helps to identify them: we discarded the sources having an extension that does not match with what expected from diffusion (i.e.\ number 12, 18, 22, 28, 32). For the remaining 27 sources, the morphological and spectral analysis we propose in the following section will help to asses their nature.

\section{Simulations and Results}
\label{sec:sim}

Among these sources, we selected one pulsar as a case study to assess the performance of the next-generation IACTs, which, thanks to their arcminute-scale angular resolution, will be capable of properly resolve these sources and distinguish TeV halos from other classes of astrophysical objects. We choose PSR J2238+5903 (number 41), a middle-aged pulsar ($\tau_c = 2.7 \times 10^4$~yr, $\dot{E} = 8.9 \times 10^{35}$ erg s$^{-1}$), located at a distance of 2.8 kpc \citep{atnf}.
This pulsar is not associated with any known SNR \cite{2025JApA...46...14G}, and was previously suggested as a potential TeV halo candidate \cite{2024ApJ...968..117Z}. 
Our modeling, presented in Sec.~\ref{sec:model}, yields an angular extension of \rev{$\theta_d(3~\rm TeV) \approx 0.68^\circ$ and $\theta_d(50~\rm TeV) \approx 0.44^\circ$, to be compared to $r_{39} = (0.51\pm0.04)^\circ$ (LHAASO-WCDA) and $(0.43\pm0.03)^\circ$ (LHAASO-KM2A).}

We generated spatial and spectral templates to compute the expected TeV emission according to the morphology described in Sec.\ \ref{sec:model}, with 60 angular bins and 30 energy bins in the 0--1$^{\circ}$ and 1--100~TeV ranges, respectively.
We used the fiducial parameters discussed in Sec.~\ref{sec:pulsar}, \rev{and we assumed that particles are accelerated from 100 GeV to 500 TeV with a power-law energy spectrum having $\alpha_e \approx 2.6$ and $\xi_{\rm CR}=58$\%, to match what observed by LHAASO-WCDA (a power-law spectrum having $\alpha_\gamma = 2.39\pm0.07$, $N_{\rm 3\,TeV} = (1.9\pm0.3)\times 10^{-13}$~\norm\ from 1--25 TeV). We finally added a cutoff at 100~TeV to the photon spectrum, to account for the spectral softening observed by LHAASO-KM2A (a power-law spectrum having $\alpha_\gamma = 3.55\pm0.10$, $N_{\rm 50\,TeV} = (2.0\pm0.1)\times 10^{-16}$~\norm\ above 25 TeV).}

Using the \texttt{Gammapy} software package~\citep{gammapy:2023, gammapy_zenodo}, we performed simulations of the $\gamma$-ray emission in the 1--100~TeV energy range with the CTAO-N (due to visibility constraints) and ASTRI Mini-Array instrument response functions (IRFs). For CTAO simulations, we assumed the $\alpha$-configuration (i.e.\ 4 Large Size Telescopes and 9 Medium Size Telescopes)
with a zenith angle of $40^\circ$ and averaged azimuth angle; prod5 v1.0 was used~\citep{ctairf}. In the case of ASTRI Mini-Array simulations, we used the full 9 telescope array IRF \citep{astri}. The observation times in the simulations were 15 hours for CTAO-N and 30 hours for ASTRI, both over an $8^\circ \times 8^\circ$ field of view. Both the astrophysical and instrumental backgrounds were simulated.

\begin{figure*}
  \centering \includegraphics[width=0.49\columnwidth]{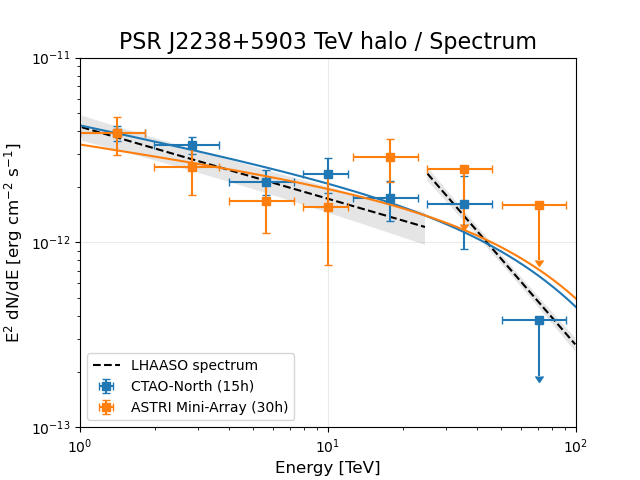}
\includegraphics[width=0.49\columnwidth]{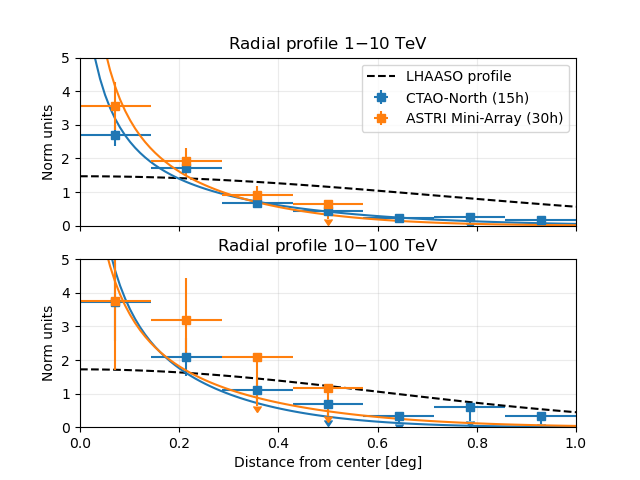}
  \caption{\footnotesize Left: Simulated $\gamma$-ray spectrum (blue: CTAO-North; orange: ASTRI Mini-Array) compared to the power-law spectra observed by LHAASO (dashed line) and to our best-fit spectra (solid lines).
  Right: Simulated radial profiles between 1--10 TeV (upper panel) and 10--100 TeV (lower panel), overlaid on the Gaussian fit reported by LHAASO (dashed line) and our best-fit model (solid lines).}
 \label{fig:sim}
\end{figure*}

Following the source simulations based on the template models outlined above, we carried out a spectral fit using the exponential-cutoff power-law model (ECPL). The spectral normalization and photon index, as well as the background normalization, were treated as free parameters. The source position and its radial extension parameters were simultaneously fitted from the spatial morphology, modelled as with a 2D Gaussian profile.
The reconstructed spectral points from CTAO-North (blue) and ASTRI Mini-Array (orange), obtained with the \texttt{FluxPointsEstimator} library, are shown in  Fig.\ \ref{fig:sim}, left panel, together with the LHAASO spectrum (dashed) and the best-fit ECPL models 
($\sqrt{\rm TS} = 12.3$, $\alpha_\gamma = 2.28\pm0.06$, $N_{\rm 3\,TeV} = (2.2\pm0.1)\times 10^{-13}$~\norm\ in the case of CTAO-North,
$\sqrt{\rm TS} = 7.9$, $\alpha_\gamma = 2.2\pm0.2$, $N_{\rm 3\,TeV} = (1.9\pm0.4)\times 10^{-13}$~\norm\ in the case of ASTRI Mini-Array).
The reconstructed radial profiles between 1--10 TeV (upper right panel of Fig.\ \ref{fig:sim}) and 10--100 TeV (lower right panel of Fig.\ \ref{fig:sim}), obtained with the \texttt{FluxProfileEstimator} library, are also shown. The best-fit $\theta_d$ are $(0.72\pm0.09)^\circ$ between 1--10 TeV, and $(0.49\pm0.11)^\circ$ between 10--100 TeV (CTAO-North), in agreement with the expected energy dependence of the diffusion radius. In the case of ASTRI Mini-Array, the two values are compatible within the error.

\section{Conclusions}

In summary, in our work we confirmed that half of the sources identified by LHAASO are in spatial coincidence with known pulsars. However, the fact that most of those are extended sources makes the emission more likely to be from sources powered by the pulsars, such as PWNe and TeV halos, or SNR associated to the pulsars.

In our analysis, we showed how 27 of the 41 \rev{pulsar associated} are in principle characterized by an extension consistent with TeV halo candidates. We discarded pulsars according to their age, observed spatial extension and flux, that we compared to the values expected for $\gamma$-rays produced by diffused leptons.

A clear-cut classification would require detailed energy-dependent morphological studies. We used PSR J2238+5903 as a case study to demonstrate that next-generation IACT facilities will be able to detect a radial profile up to angular separations of $\sim1^{\circ}$ from the center of the source in only 15 hr (CTAO-North) and 30 hr (ASTRI Mini-Array) of exposure. Not only will this capability provide a deeper understanding of these sources and the other halos, but also essential constraints on the particle acceleration and propagation mechanisms.

\section*{Acknowledgments}
\small
This work was conducted in the context of the ASTRI and CTAO Projects.
We gratefully acknowledge support from the people, agencies, and organisations listed here: \href{http://www.astri.inaf.it/en/library/}{http://www.astri.inaf.it/en/library/}.
We gratefully acknowledge financial support from the agencies and organizations listed at \href{http://www.ctao.org/for-scientists/library/acknowledgments/}{http://www.ctao.org/for-scientists/library/acknowledgments/}. This research has made use of the CTA instrument response functions provided by the CTA Consortium and Observatory, see \href{http://www.cta-observatory.org/science/ctao-performance}{http://www.cta-observatory.org/science/ctao-performance} for more details.
This paper went through the internal ASTRI and CTAO review processes.
We thank Elena Amato for her helpful comments.
M.R.\ and A.B.\ acknowledge the NRRP - funded by the European Union - NextGenerationEU (CUP C53C22000430006).

\small
\bibliographystyle{JHEP}
\bibliography{bibliography.bib}



\end{document}